\documentclass[a4paper,fleqn,usenatbib]{mnras}
\usepackage[T1]{fontenc}
\usepackage{ae,aecompl}

\usepackage[normalem]{ulem}
\usepackage{amsmath}
\usepackage{graphicx}	
\usepackage{lscape}
\usepackage{indentfirst}
\usepackage{enumitem}

\usepackage{amssymb}
\usepackage{color}
\usepackage[flushleft]{threeparttable}

\newcommand {\bc}{\begin {center}}
\newcommand {\ec}{\end {center}}
\newcommand {\be}{\begin {equation}}
\newcommand {\ee}{\end {equation}}
\newcommand {\beq}{\begin {eqnarray}}
\newcommand {\eeq}{\end {eqnarray}}

\renewcommand{\d}{{\rm d}}
\newcommand{\ampl}{a}

\title[No beaming in pulsating ULXs]
{Pulsating ULXs: large pulsed fraction excludes strong beaming}
\author[A. A.~Mushtukov et al.] 
{Alexander~A.~Mushtukov,$^{1,2}$\thanks{E-mail: al.mushtukov@gmail.com (AAM)}  
Simon Portegies Zwart,$^{1}$
Sergey~S.~Tsygankov,$^{3,2}$
\newauthor
Dmitrij~I.~Nagirner$^{4}$
and
Juri Poutanen$^{3,2,5}$\\ 
$^1$ Leiden Observatory, Leiden University, NL-2300RA Leiden, The Netherlands \\
$^2$ Space Research Institute of the Russian Academy of Sciences, Profsoyuznaya Str. 84/32, Moscow 117997, Russia \\
$^3$ Department of Physics and Astronomy,  FI-20014 University of Turku, Finland \\  
$^4$ Sobolev Astronomical Institute, Saint Petersburg State University, Saint-Petersburg 198504, Russia\\
$^5$ Nordita, KTH Royal Institute of Technology and Stockholm University, Roslagstullsbacken 23, SE-10691 Stockholm, Sweden
} 

\pubyear{2020}

\begin{document}
\label{firstpage}
\pagerange{\pageref{firstpage}--\pageref{lastpage}}
\maketitle

\begin{abstract}    
The recent discovery of pulsating ultra-luminous X-ray sources (ULXs) shows that the apparent luminosity of accreting neutron stars (NSs) can exceed the Eddington luminosity by a factor of hundreds.
The relation between the actual and apparent luminosity is a key ingredient in theoretical models of ULXs but it is still under debate.
A typical feature of the discovered pulsating ULXs is a large pulsed fraction (PF).
Using Monte Carlo simulations, we consider a simple geometry of accretion flow and test the possibility of simultaneous presence of a large luminosity amplification due the geometrical beaming and a high PF.
We argue that these factors largely exclude each other and only a negligible fraction of strongly beamed ULX pulsars can show PF above 10 per cent.
Discrepancy between this conclusion and current observations indicate that pulsating ULXs are not strongly beamed and their apparent luminosity is close to the actual one.
\end{abstract}



\section{Introduction}
\label{sec:Intro}

The recent discovery of pulsating ultra-luminous X-ray sources (ULXs, see e.g., \citealt{2014Natur.514..202B,2016ApJ...831L..14F,2017Sci...355..817I,2018MNRAS.476L..45C,2020ApJ...895...60R}) implies that a significant fraction of ULXs is represented by accreting strongly magnetised neutron stars (NSs).
All discovered pulsating ULXs show pulsations episodically when the pulsed fraction (PF) is large and detected to be above 10--20 per cent. 
Search of pulsations requires a large amount of counts. 
Only $\sim$15 out of $\sim$300 known ULXs \citep{2019MNRAS.483.5554E}  observed by XMM-Newton provide the statistics sufficient for detection of pulsations, and $\sim$25 per cents of them are proven to be accreting NSs (see detailed discussion in \citealt{2020ApJ...895...60R}).
Therefore, we can speculate that a significant PF is a typical feature of pulsating ULXs.

\begin{figure*}
\centering 
\includegraphics[width=15.cm]{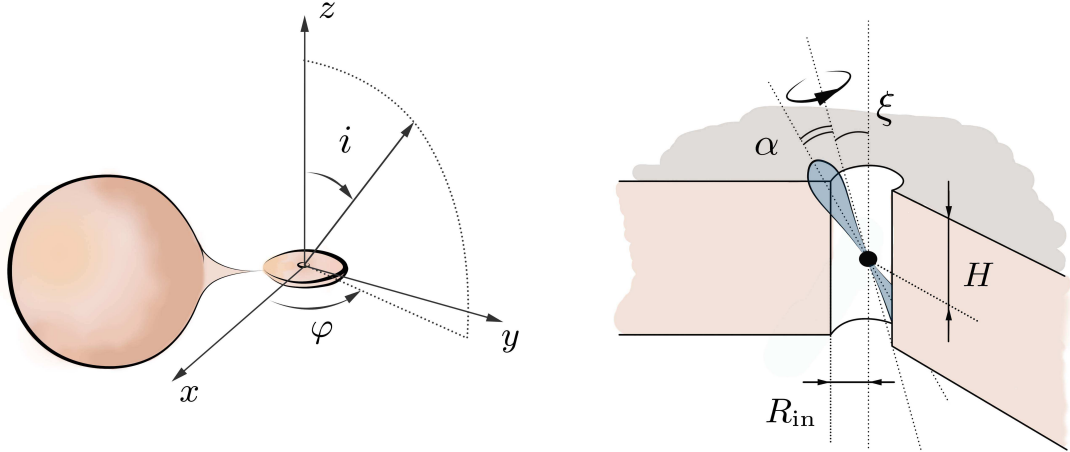} 	
\caption{Schematic illustration of the considered geometry. 
The accretion flow from the companion star forms an accretion disc around the central object. 
The accretion disc plane is close to the orbital plane of the binary system.
The accretion flow in the vicinity of the compact object is geometrically thick. 
}
\label{pic:scheme}
\end{figure*}

The typical magnetic field strength at the surface of accreting NSs in pulsating ULXs is still under debate.
Different theories provide predictions, which cover a few orders of magnitude from $10^{10}$ up to $10^{14}$~G.
It is also unclear what the relation between the actual $L$ and the apparent $L_{\rm app}$ luminosity is.
The extreme mass accretion rates imply geometrically thick radiation pressure dominated accretion discs and probably strong outflows from the discs \citep{1973A&A....24..337S,1999AstL...25..508L,2007MNRAS.377.1187P,2017ApJ...845L...9T},
{which were already detected in a few sources (see e.g., \citealt{2016Natur.533...64P,2018MNRAS.473.5680K,2018MNRAS.479.3978K,2020MNRAS.491.5702P})}.
Thus, certain beaming of X-ray luminosity is expected.
Beaming affects the apparent luminosity $L_{\rm app}$ amplifying it with respect to the actual accretion luminosity $L$: 
\beq
L_{\rm app}=\ampl L.
\eeq
The amplification factor $\ampl$ in ULX pulsars is assumed to be close to unity by some authors \citep[see, e.g.,][]{2015MNRAS.449.2144D,2015MNRAS.448L..40E,2015MNRAS.454.2539M,2017MNRAS.467.1202M,2019MNRAS.484..687M,2019A&A...626A..18C} and suggested to be large $\ampl \gtrsim 20$ by other authors  \citep[see, e.g.,][]{2017MNRAS.468L..59K,2017MNRAS.470L..69M,2019MNRAS.485.3588K,2020MNRAS.494.3611K}.
In particular, \citet{2009MNRAS.393L..41K} argues that advection in accretion disc around magnetised NS is negligible and all mass accretion rate above the local Eddington limit is lost due to the outflow.
As a result, an accreting NS is located inside a well-like tunnel, whose radius is close to the inner radius $R_{\rm in}$ of the accretion disc and the geometrical depth $H$ is determined by the outflow and is considered a parameter in this paper (see Fig.\,\ref{pic:scheme}).

There are two problematic points for theoretical models which require a substantial amplification factor at high apparent luminosities of X-ray pulsars {($\ampl\sim 20$ for M82~X-2, NGC~7793~P13 and NGC~300~ULX1, and $\sim$100 for NGC~5907~ULX1, see Table~1 in \citealt{2020MNRAS.494.3611K})}.
The first one is a large observed PF, which is expected to be reduced in systems with strong beaming.
Recently, however, it was proposed that strong pulsations in ULXs appear in the case of a lucky combination between geometrical parameters of the accreting system \citep{2020MNRAS.494.3611K}.
Another problematic point is related to bright transient X-ray pulsars, which do not show any evidence of beaming at high luminosity states \citep{2017A&A...605A..39T,2020MNRAS.491.1857D,2020MNRAS.495.2664C}.
Note, that according to \citealt{2018A&A...620L..12V,2019MNRAS.488.5225V,2020MNRAS.491.4949V} there is no need for strong beaming even in some pulsating ULXs.
Additionally, strong beaming in ULXs implies a large number of unbeamed and even diluted sources.

In this paper, we test numerically the possibility of a simultaneous presence of strong beaming and large PF in pulsating ULX using a simplified geometry of accretion flow similar to that proposed by \citet{2020MNRAS.494.3611K}.

\section{Model}
\label{sec:NumModel}

We use a geometry where the NS is located inside the cylindrical cavity of radius $R_{\rm in}$ and height $2H$ (see Fig.\,\ref{pic:scheme}).
Two geometrical parameters determine the {observed properties of a rotating}  magnetized NS: 
the angle between the orbital axis and the NS spin axis $\xi$, 
and the magnetic obliquity $\alpha$, i.e. angle between the rotational axis and the magnetic axis of a NS.
The $z$-axis is taken to be aligned with rotational axis of the accretion flow, 
the NS spin is in the $x-z$ plane, 
and the $y$-axis completes the right-handed coordinate system. 
The observer's position is specified by the inclination $i$ and azimuthal angle $\varphi$.
We assume that the magnetic field of a NS is dominated by dipole component, and an accreting NS produces two beams of X-rays directed along the magnetic field axis.
{The angular distribution of the luminosity is assumed to follow the law:} 
\beq\label{eq:beam}
\frac{\d L(\theta)}{\d\cos\theta}\propto \cos^n\theta,
\eeq
where $\theta\in[0;\pi/2]$ is the angle between the magnetic axis and photon momentum. 
Photons emitted by accreting NS leave the system immediately if their momenta are directed within a narrow cone along the accretion flow axis.
Otherwise, photons are reflected multiple times by the walls of the funnel before they leave the cavity.

Using Monte Carlo simulations, we can trace the history of each photon (see Appendix \ref{App:Code}).
Our code is designed under the following set of assumptions: 
\begin{itemize}
\item  the NS magnetic field is dominated by a dipole component and photons are emitted in the vicinity of two magnetic poles at the stellar surface {according to the law \eqref{eq:beam};} 
\item reflection of a photon by the walls of accretion funnel is calculated under the assumption of multiple conservative, isotropic and coherent scattering in the semi-infinite medium;
\item  the light travel time inside the accretion funnel is assumed to be much shorter than the NS spin period (validity of this assumption is discussed in more detail in Sect. \ref{sec:time}).
\end{itemize}

Tracing a history of $N$ photons emitted by a NS at certain geometrical configuration ($N=10^8$ in the simulations represented in this paper), we get the distribution of photons over the final directions. 
This distribution is related to the photon energy flux, which is detected by a distant observer at the specific direction from a system.
Rotation of a NS results in variability of the observed X-ray energy flux $F_{\rm obs}$ during the spin period $P$.
Fixing two angles,  $\alpha$ and $\xi$, that describe rotation of a magnetised NS with respect to the accretion flow (see Fig.\,\ref{pic:scheme}, right), we produce pulse profiles $F_{\rm obs}(i,\varphi; t)$ detected by distant observers situated at different inclinations $i$ and azimuths $\varphi$.  
From the pulse profile we determine the average X-ray flux 
\beq\label{eq:F_ave}
F_{\rm ave}(i,\varphi)=\frac{1}{P}\int_{0}^{P} F_{\rm obs}(i,\varphi; t) \ \d t\  , 
\eeq
and the PF
\beq 
\mbox{PF} (i,\varphi) = \frac{F_{\rm max}(i,\varphi)-F_{\rm min}(i,\varphi)}{F_{\rm max}(i,\varphi)+F_{\rm min}(i,\varphi)},
\eeq
where $F_{\rm min}$ and $F_{\rm max}$ are minimal and maximal flux during the pulsation period.
Both $F_{\rm ave}$ and PF depend on the direction to the observer $(i,\varphi)$. 
The average flux $F_{\rm ave}(i,\varphi)$ is used to get the luminosity amplification factor 
\beq
a(i,\varphi) = \frac{F_{\rm ave}(i,\varphi)}
{(4\pi)^{-1}\int_{4\pi} F_{\rm ave}(i,\varphi) \ \d\Omega} ,
\eeq 
which connects the actual and apparent luminosity.
 
A system described by a given set of parameters can be detected by an observer with a different combination of the amplification factor $\ampl$ and PF depending on the orientation of the observer in the reference frame of a system.
The distribution of the amplification factor and PF over the directions is related to the probability of ULX to be detected with a particular amplification factor and PF.

{The distribution over the amplification factor $a$ and PF depends on the geometrical parameters $\xi$ and $\alpha$ (see Fig.\,\ref{pic:scheme}).}
Assuming a specific distribution of accreting NSs over these angles, we can get a general distribution of systems over the amplification factor $\ampl$ and PF for a given ratio $H/R_{\rm in}$ and given parameter $n$.
Using this distribution, we can test the hypothesis that apparent luminosity of pulsed ULXs is significantly larger than the actual accretion luminosity.

{Calculating the distribution functions of ULXs over the amplification factors and PF, we assumed random orientation of rotating NS with respect to the accretion flow, i.e. the angles $\alpha$ and $\xi$ (see Fig.\,\ref{pic:scheme}) take random values in the interval $[0;\pi]$ (the distribution functions based of random $\cos\alpha$ and $\cos\xi$ are very similar to those calculated under assumptions of random $\alpha$ and $\xi$).
However, the interaction of a NS with accretion flow affects both orientations of rotation axis \citep{2014EPJWC..6401001L} and magnetic dipole axis.
If there is no precession of accretion disc, the spin of a NS is aligned with the axis of accretion disc in the equilibrium.
The relaxation time to the equilibrium depends on the magnetic dipole moment of a NS $\mu$ and mass accretion rate onto the NS surface.
In the case of Eddington mass accretion rate the relaxation time can be estimated as 
$t_{\rm rel}\approx 300 I_{45}m^{1/3}\mu_{30}^{-4/3}$ years \citep{1982SvA....26...54L}, where $I_{45}$ is the NS momentum of inertia in units of $10^{45}\,{\rm g\,cm^2}$ and $\mu_{30}$ is the NS dipole magnetic moment in units of $10^{30}\,{\rm G\,cm^3}$.
As a result, we would expect that the rotational axis is aligned with the disc axis in ULX, hosting NS of an extremely strong magnetic field. In contrast, the orientation of the rotational axis of weakly magnetised NS can be far from the equilibrium and oriented randomly.
Fixing the rotational axis to be aligned with the accretion disc axis ($\xi=0$), we get the distributions, which are only slightly different from the ones with random orientation of the rotational axis. 
}

{
Note that assuming conservative scatterings in the accretion cavity walls, we neglect the possibility of true absorption of X-ray photons.
True absorption of X-rays in the cavity contributes to a loss of information about pulsations and complicates even more detection of pulses. 
}

\section{Numerical results}
\label{sec:NumResults}

\subsection{Luminosity amplification}

\begin{figure}
\centering 
\includegraphics[width=8.0cm]{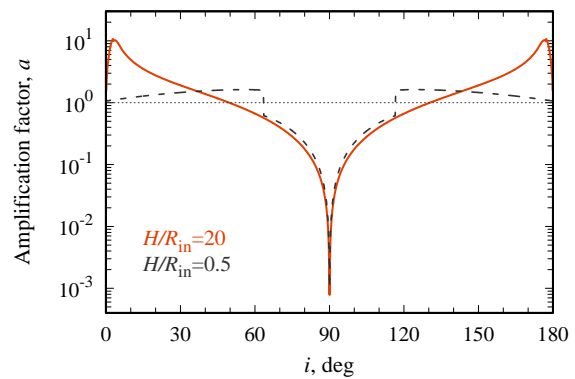} 	
\caption{
Dependence of the amplification factor on the inclination angle $i$ (see Fig.\,\ref{pic:scheme}). 
Different curves are given for different ratios $H/R_{\rm in}=20$ (red solid), 0.5 (dashed black).
The larger the ratio $H/R_{\rm in}$, the stronger the beaming along the accretion disc axis.
The flux in the plane of the accretion flow ($i=90^{\circ}$) is suppressed due to the eclipsing of the central source by accretion disc.
}
\label{pic:sc_beaming}
\end{figure}

\begin{figure}
\centering 
\includegraphics[width=8.0cm]{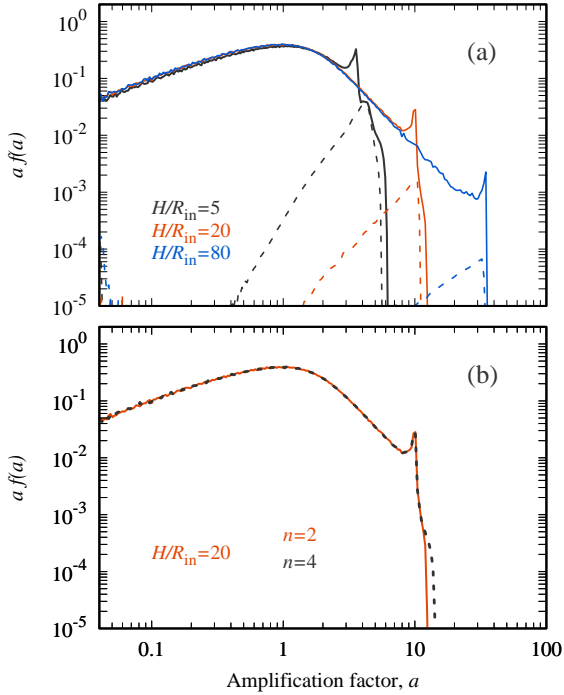}
\caption{
Pulsar distribution over the luminosity amplification factor $\ampl$. 
{
(a) Different solid curves show the distributions calculated for different ratios $H/R_{\rm in}=5$ (black),  20 (red), 80 (blue).
The larger the $H/R_{\rm in}$ ratio, the larger the maximal possible luminosity amplification.
We assumed here $n=2$.
The dashed curves represent the fractions of the distributions with PF above 10 per cent.
We see that only a small fraction of sources show high  PF.
(b) The distributions given by red solid and black dashed lines are calculated for parameter $n=2$ and $4$ at fixed $H/R_{\rm in}=20$.
Parameter $n$ affects the distributions only slightly.
}
}
\label{pic:sc_LPF_f_L}
\end{figure}

\begin{figure}
\centering 
\includegraphics[width=8.5cm]{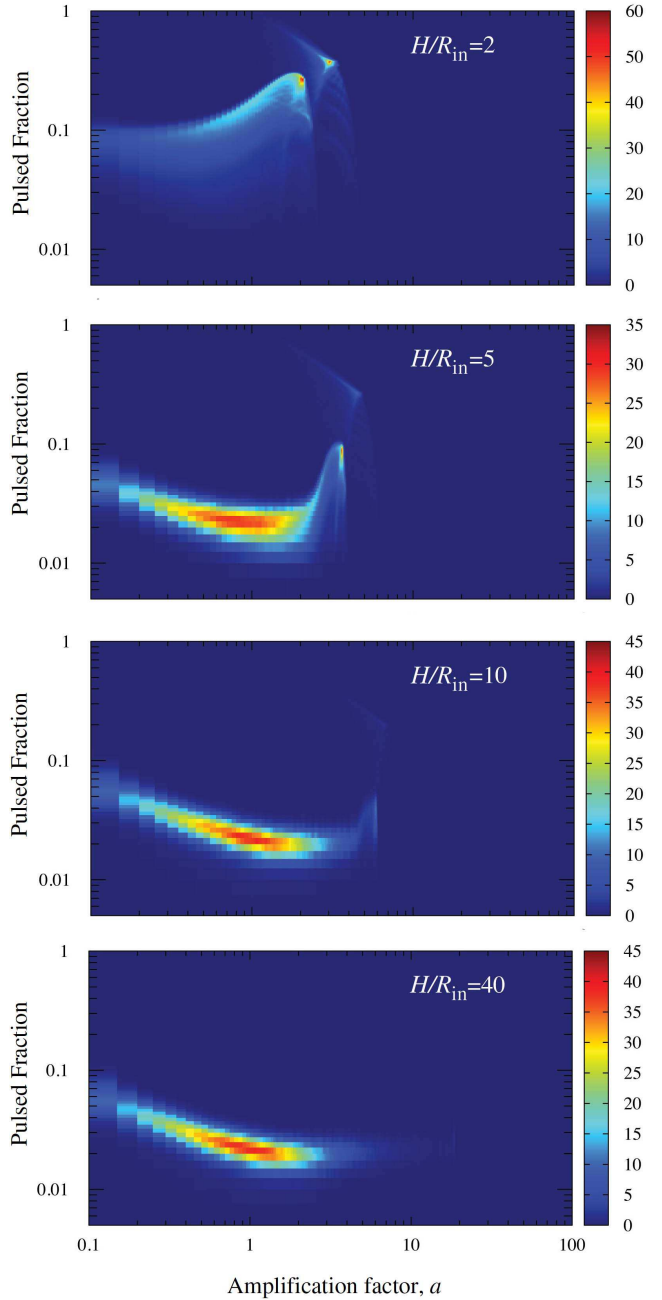} 	
\caption{
Two-dimensional distribution of sources over the luminosity amplification factor $\ampl$ and the PF.
Different panels represent the distribution of the case of different ratios $H/R_{\rm in}=2, 5, 10, 40$ (from top to bottom).
The larger ratio $H/R_{\rm in}$ leads to larger maximal amplification factors and smaller typical PF.
}
\label{pic:sc_LPF_2d}
\end{figure}

The apparent luminosity of ULX pulsars derived from the flux averaged over the pulsation period $F_{\rm ave}$ (see equation \ref{eq:F_ave}) is different than the actual luminosity even in the case of geometrically thin discs.
It happens due to the natural beaming of rotating non-isotropic sources. 
The beaming due to the geometry of the accretion flow tends to be stronger along the axis of accretion disc (see Fig.\,\ref{pic:sc_beaming}). 
The larger the ratio of $H/R_{\rm in}$, the larger the maximal amplification factor.
The average flux for observers looking at the system from the directions of large $i$ (nearly equatorial plane) is small, which results in a low apparent luminosity.
The probability of a certain inclination $i$ is $\propto\sin i$.
Therefore, a significant fraction of systems with a large $H/R_{\rm in}$ ratio show low apparent luminosity even in the case of a large maximal amplification (see Fig.\,\ref{pic:sc_LPF_f_L}).
Thus, the existence of strongly beamed sources in observations implies that there are many more sources with a low apparent luminosity.

\subsection{Beaming versus pulsed fraction}

The increase of beaming at large $H/R_{\rm in}$ is accompanied by a decrease of the PF (see Fig.\,\ref{pic:sc_LPF_2d}).
The absolute majority of sources show PF below 10 per cent at $H/R_{\rm in}>5$ already.
The decrease of a PF at a large $H/R_{\rm in}$ ratio is natural because photons experience multiple scatterings, which result in a loss of information about initial photon momentum.
A certain fraction of sources still shows both strong luminosity amplification and large PF, but their population among all sources is negligibly small (see Fig.\,\ref{pic:sc_LPF_f_L}): for the case of $H/R_{\rm in}=5$, only 2 per cents of sources show PF above 10 per cents, while for the case of $H/R_{\rm in}=80$ the fraction of sources with PF above 10 per cents drops below 0.1 per cent.
{In general, less than 1 per cent of strongly amplified sources (with the amplification factor $a>10$) show the PF above 10 per cent.
}
All the sources with simultaneously large amplification factor and PF are visible for the observers looking at the system almost along the axis of the accretion disc.

The distribution of the sources over the amplification factor and the PF depend on the sharpness of the initial beam. 
The initial beam can be complicated and affected by several factors like geometry of emitting region \citep{1976MNRAS.175..395B}, the gravitational bending of X-ray photons \citep{2018MNRAS.474.5425M,2020PASJ...72...34I}, and photon reprocessing by the accretion flow between the accretion disc inner radius and the NS surface \citep{1976SvAL....2..111S,2017MNRAS.467.1202M}.
The sharpness of the initial beam from a NS surface is described in our simulations by parameter $n$ (see equation \ref{eq:beam}).
{At large $H/R_{\rm in}$ ratios, parameter $n$ only slightly affects the distribution of sources over the amplification factor (see Fig.\,\ref{pic:sc_LPF_f_L}b).}

\begin{figure}
\centering 
\includegraphics[width=8.7cm]{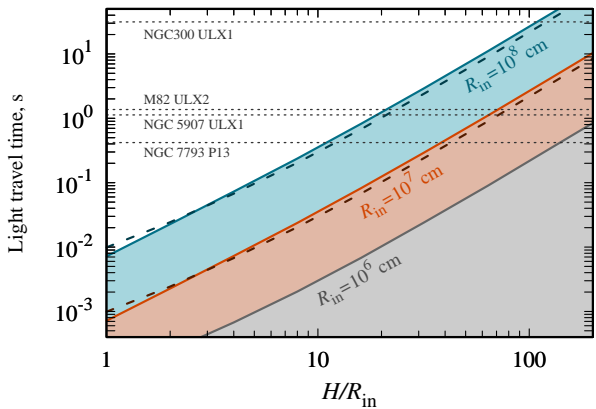} 	
\caption{
The average light travel time in a system (solid lines) and the standard deviation of the light travel time (dashed lines) as a function of $H/R_{\rm in}$.
Blue, red and grey lines are given for the case of inner disc radii $R_{\rm in}=10^8,\,10^7$ and $10^6$\,cm, respectively.
If the spin period of a NS is smaller than the light travel time (colored regions below the solid lines), pulsations from a ULX are undetectable for a given geometry of accretion flow.
Horizontal dotted lines represent the observed pulsation periods of four ULX pulsars: 
NGC 300 ULX1 \citep{2018MNRAS.476L..45C}, 
M82 X-2 \citep{2014Natur.514..202B,2020ApJ...891...44B}, 
NGC 5907 ULX1 \citep{2017Sci...355..817I},
NGC 7793 P13 \citep{2016ApJ...831L..14F,2017MNRAS.466L..48I}.
}
\label{pic:sc_King_ave}
\end{figure}

\subsection{Light travel time}
\label{sec:time}

The procedure of pulse shape construction in our numerical model assumes that the spin period of a NS is longer than the typical light travel time inside the accretion funnel (see Fig.\,\ref{pic:scheme}, right).
If it is not a case and light travel time is larger than the spin period, the pulsations are expected to vanish from the signal.

The geometrical size of a system at given ratio $H/R_{\rm in}$ is determined by the inner radius of accretion flow, which is expected to be close to the magnetospheric radius
\beq 
\label{eq:Rm}
R_{\rm m}\approx 1.2\times 10^{8}\,\Lambda {B}_{12}^{4/7}\dot{M}_{18}^{-2/7}
m^{-1/4}R_6^{12/7}\ \mbox{cm},
\eeq 
where $\Lambda$ is a coefficient expected to be close to unity for the case of radiation-pressure dominated accretion discs \citep{2019A&A...626A..18C},
$B_{12}$ is the magnetic field strength at the NS surface in units of $10^{12}$~G,
$\dot{M}_{18}$ is the mass accretion rate at the inner radius of the disc in units of $10^{18}$\,g\,s$^{-1}$, and $R_6$ is the NS radius in units of $10^6$\,cm.

Tracing photon history and neglecting the time of a photon reprocessing by the walls of accretion funnel, we get an upper limit on the average light travel time inside the system $\langle t\rangle$ and its standard deviation $\sigma(t)$ (see Fig.\,\ref{pic:sc_King_ave}).
Both of them tend to increase at $H/R_{\rm in}\gg 1$: $\langle t\rangle, \sigma(t)\propto (H/R_{\rm in})^2$ .
Therefore, detection of high-frequency pulsation is expected to be hard or even impossible. 
The detection of pulsations of certain frequency from a ULX naturally puts an upper limit on the $H/R_{\rm in}$ ratio and maximal possible amplification of the apparent luminosity.

\section{Summary}
\label{sec:Sum}

In this work, we tested the hypothesis that the apparent luminosity of detected ULX pulsars is strongly amplified relative to the actual luminosity by the collimating geometry of the accretion flow  \citep[see, e.g.,][]{2017MNRAS.468L..59K,2019MNRAS.485.3588K,2020MNRAS.494.3611K}.
To do that we used a simplified model of the accretion flow geometry (see Fig.\,\ref{pic:scheme}) representing the collimating wind as a cylinder.
We performed Monte Carlo simulations tracing photon history from their emission in the vicinity of a NS surface till their escape from the system. 
We computed the luminosity amplification factor $\ampl$ and the PF as a function of the observer's orientation with respect to ULX pulsar for a given geometry of the accretion flow.

Our simulations show that large PF detected in ULX pulsars  excludes strong amplification of the accretion luminosity.
Only a tiny fraction of systems show simultaneously a large (above 10 per cent) PF and significant luminosity amplification due to the geometrical beaming (see Fig.\,\ref{pic:sc_LPF_2d}).
The majority of sources do not show pulsations with PF$>10$ per cent (see Fig.\,\ref{pic:sc_LPF_f_L}).
The considered geometry erases pulsations from the signal not only due to redistribution of X-ray photons over directions, but also because of possibly large light travel time inside the accretion funnel (see Fig.\,\ref{pic:sc_King_ave}).
In particular, the considered geometry makes it impossible to detect pulsed ULX with small spin periods.
{
The transient nature of pulsation in ULXs is beyond the scope of this paper.
The disappearance of pulsations can be related to the processes in close proximity to the NS.
In the case of geometrical beaming, strong amplification of the apparent luminosity excludes any strong pulsations.
In that sense, our conclusions are applicable to the peak PF observed in ULX pulsars. 
}

We conclude that the theoretical models assuming strong geometrical beaming are inconsistent with observations, where pulsations  with a large PF are already detected in a quarter of ULXs having sufficient statistics \citep{2020ApJ...895...60R}.
The resent results of population synthesis models also show that there is no need for strong beaming in ULX pulsars \citep{2020arXiv201003488K}.
The exclusion of strong beaming argues in favour of super-Eddington accretion onto magnetized NSs as a reason for high apparent luminosity of ULX-pulsars.

\section*{Acknowledgements}

This work was supported by the Netherlands Organization for Scientific Research Veni Fellowship (AAM) and the grant 14.W03.31.0021 of the Ministry of Science and Higher Education of the Russian Federation.
We are grateful to Paolo Esposito, Valery Suleimanov and an anonymous referee for useful comments and discussions.

\section*{Data availability}

The calculations presented in this paper were performed using a private code developed and owned by the corresponding author. All the data appearing in the figures are available upon request. 

\bibliographystyle{mnras}
\bibliography{allbib}

\appendix

\section{Description of Monte Carlo code}
\label{App:Code}

{
We use the Monte Carlo code written in Fortran 95 to trace the history of X-ray photons emitted from two polar regions at the surface of accreting NS.
The main parameters of the simulations are $H/R_{\rm in}$ ratio, the inner radius of accretion flow $R_{\rm in}$, which scales the light travel time in a system, parameter $n$ determining the sharpness of a beam pattern, and angles $\alpha$ and $\xi$, which determine together with a phase angle the orientation of a NS in respect to the accretion flow (see right panel Fig.\,\ref{pic:scheme} ). 
As a result of numerical simulation, we get photons distribution over the final directions, which are given by angles $i$ and $\varphi$ (see left panel Fig.\,\ref{pic:scheme}). 
}

{
There are a few steps in Monte Carlo simulations:
}

\begin{enumerate}[label= \arabic*)]
\item 
{
We generate a new photon at the NS surface.
The direction of the emitted photon is obtained out of three random numbers $X_i\in [0;1]$. 
The first random number $X_1$ determines which one of two polar regions emits the photon.
The second $X_2$ and the third $X_3$ random numbers determine the polar $\theta_{\rm B}$ and the azimuthal $\varphi_{\rm B}$ angles of the photon momentum in the reference frame, where the $z$-axis is aligned with the magnetic axis of a NS: 
\beq 
\theta_{\rm B}&=&\arccos\left[(-1)^{{\rm sgn}(X_1-0.5)}X_2^{1/(1+n)} \right], \\
\varphi_{\rm B}&=&2\pi X_3.
\eeq 
The simulations presented in this paper are based on tracking of $10^8$ photons. 
}

\item 
{As soon as we know the direction of photon motion in the reference frame of a NS and orientation of a NS in respect to the accretion flow (which is determined by angles $\alpha$ and $\xi$), we get the direction of photon motion in the reference frame of the accretion flow, where the $z$-axis is aligned with the rotational axis of the disc. 
The photon moving along straight trajectories either leaves the cylindrical cavity or crosses the wall of the cavity.
If the photon leaves the cavity, we account for it in the distribution of the final flux over the directions.
If the photon crosses the wall, we simulate the reprocessing of the photon (see Step 3).
At this step, we track the light travel time of a photon inside the accretion cavity.
}

\item
{
We simulate the reprocessing of a photon, which crosses the wall of accretion cavity.
The reprocessing is calculated under the assumption of conservative isotropic scattering in a semi-infinite medium. 
The process is considered as a series of scatterings. 
The free path of a photon between the scattering events inside the accretion flow is calculated as
\beq 
\Delta \tau = -\ln X_4,
\eeq 
where $X_4$ is a random number, and the free path $\Delta\tau$ is given in units of optical thickness, i.e., dimensionless.
Tracking the direction of photon momentum and the free path, we know the optical depth of each scattering event and displacement of a photon in respect to the edge of a cavity. 
The direction of photon momentum after the isotropic scattering is given by 
\beq
\theta_{\rm f}=\arccos(1-2X_5),\quad \varphi_{\rm f}=2\pi X_6.
\eeq
We track the series of scattering until the photon leaves the semi-infinite medium.
When it happens, the current direction of photon momentum is taking for a new direction of photon motion inside the accretion cavity, and calculations return to Step 2.
Simulating the reprocessing of a photon by semi-infinite medium, we neglect the delays related to this process, assuming that they are much smaller than photon travel time inside the accretion cavity.
As a result, the light travel time obtained in our simulations gives a lower limit for the actual light travel time.
}
\end{enumerate}

{
The distributions of the photons over the final direction obtained as a result of the simulations are used to construct the pulse profiles and the amplification factors  (see Section \ref{sec:NumModel}).
The latter can be averaged over the distributions of $\alpha$ and $\xi$.
Our numerical code was tested against the geometries, where the ratio $H/R_{\rm in}$ is relatively small. 
In particular, the case of $H/R_{\rm in}=0.5$ illustrated in Fig.\,\ref{pic:sc_beaming} indicates the eclipses of the central source by the accretion disc in the equatorial plane of a system.
The eclipse is detected for the angles within $\sim 26^\circ$ above and below the equatorial plane, which agrees with the expectations: $\arctan(0.5)\sim 26^\circ$.
}


\bsp	
\label{lastpage}
\end{document}